\documentstyle{aa}
\input{psfig.sty}

\newcommand{\PSR}{PSR~B1821$-$24 }

\def\fdeg{\hbox{$\,.\!\!^{\circ}$}}

\begin{document}

\title{Spectral and timing properties of the X-ray emission from the 
millisecond pulsar PSR~B1821$-$24}

\author{T. Mineo\inst{1}, G. Cusumano\inst{1}, E. Massaro\inst{2,3}, 
W. Becker\inst{4}, L. Nicastro\inst{1}}

\institute{Istituto di Astrofisica Spaziale e Fisica
Cosmica, INAF-CNR, Sezione di Palermo, Via Ugo La Malfa 153, 
I-90146, Palermo, Italy \and
Dipartimento di Fisica, Universit\'a La Sapienza, Piazzale A. Moro 2,
I-00185, Roma, Italy \and
Istituto di Astrofisica Spaziale e Fisica Cosmica, INAF-CNR,
Sezione di Roma, Via Fosso del Cavaliere,
I-00113, Roma, Italy \and
Max-Planck-Institut f\"ur Extraterrestrische Physik, D-85740, 
Garching-bei-M\"unchen, Germany}

\offprints{T. Mineo: mineo@pa.iasf.cnr.it}

\date{Received: 4 February 2004; accepted: 27 April 2004}

\titlerunning{PSR~B1821$-$24}

\authorrunning{T. Mineo et al.}

\abstract{We report results on the timing and spectral analysis of 
observations of the millisecond pulsar \PSR with RXTE, BeppoSAX and 
Chandra. 
The X-ray light curve is characterized by two narrow peaks at a 
phase distance of $0.452\pm0.002$.
The average pulsed emission, over the range 1.6--20 keV, is well represented 
by a single power law with a photon index $\alpha=1.30_{+0.05}^{-0.02}$ 
and unabsorbed (2--10 keV) pulsed X-ray flux of $3.9\times10^{-13}$ 
erg cm$^{-2}$ s$^{-1}$.
We searched for a possible bunching of X-ray photons to verify if
the X ray emission has a time structure similar to that of giant
pulses and found a negative result. 
 }

\maketitle

\section{Introduction}

PSR~B1821$-$24 is an isolated millisecond radio pulsar discovered
in the core of the globular cluster M28 (Lyne et al. 1987). It is
one of the most powerful (spin-down luminosity $2.2\times10^{36}$ 
erg s$^{-1}$) and youngest (spin-down age $3\times10^{7}$ yr) 
millisecond pulsars (MSPs).
This solitary MSP has a rotation period of 3.05 ms and a period 
derivative of $\dot{P}=1.61\times 10^{-18}$ s s$^{-1}$.
The inferred magnitude of the dipolar field at the magnetic pole
is $4.5\times 10^{9}$ G, the highest among the known MSPs.
The distance of the source, assumed equal to the globular cluster 
one, ranges from  $5.1\pm0.5$ kpc (Rees \& Cudworth 1991) to 
5.7 kpc (Harris 1996). In analogy with Becker et al. (2003) we use 
5.5 kpc in our calculations.

Marginal detection of X-ray pulsed emission was reported by Danner,
Kulkarni \& Thorsett (1994) from  a ROSAT (PSPC) observation from 
which they inferred a luminosity of $2.9\times10^{33}$ erg s$^{-1}$ 
in the energy band 0.1--2.4 keV.  ASCA observation of \PSR (Saito 
et al. 1997) evidenced a double peak profile with different shapes and a 
phase separation of $0.44\pm0.01$.\\
Studies of the absolute phase in order to establish the phase lag
of the X-ray emission with respect to the radio pulses have been
performed with RXTE and Chandra observations (Rots et al. 1998; 
Rutledge et al. 2003).
The absolute phase of the higher and narrower of the two
X-ray  peaks significantly 
lags the corresponding radio feature, named P1 by Backer \& Sallmen 
(1997), of $60\pm20$ $\mu$s.
\\
The total pulsed spectrum was modelled with a single power law 
of spectral index $1.30\pm0.07$ for the ASCA/GIS (0.8--10 keV) data, 
$1.15\pm0.02$ for the RXTE/PCA data (2--20 keV) and $1.13\pm0.02$ for 
the joint ASCA-RXTE data (Kuiper et al. 2003).

Radio giant pulses have been observed  from \PSR  with intensities
that reach 80 times the average pulse energy. The brightest pulses
are 
concentrated in a narrow phase window ($\sim$0.07), 
as derived from the average pulsed profile, that lags P1 by 80 $\mu$s  
and so it is coincident with the main X-ray peak (Romani \& Johnston 
2001).

Recently, the high spatial resolution of Chandra allowed us to
resolve 12 sources within the core of M28 where the pulsar is located 
(Becker et al. 2003; Rutledge et al. 2003). 
This made it possible to measure the unconfused phase averaged
X-ray spectrum and the level of the unpulsed emission. The total
(pulsed plus unpulsed) X-ray spectrum was modeled by a single
power law with a photon index of $1.2^{+0.15}_{-0.13}$ and an 
 unabsorbed (0.5--8 keV) flux of $3.5\times10^{-13}$ erg cm$^{-2}$ 
s$^{-1}$ with about 15\% assigned to the non-pulsed emission
(Rutledge et al. 2003). Moreover, the residuals in the Chandra
spectrum show a hint of a feature at 3.3 keV, that  Becker et al.
(2003) fitted with a Gaussian. This line, having a width of 0.8 keV
and a flux of $6\times10^{-6}$ photons cm$^{-2}$ s$^{-1}$, was
identified as an electron cyclotron line formed in a
magnetic field of $3\times10^{11}$ G. The value, much higher than
that derived from the dipole radiation, was explained either with 
a possible presence of multipolar components or with a strong 
displacement of the magnetic dipole center.

In this paper we present the timing and spectral analysis of one
BeppoSAX and three RXTE observations of the pulsar covering the
energy range 1.6--20 keV, including, for the spectral analysis,
data from a Chandra/ACIS observation.

\section{Observation and data reduction}

RXTE observed the source three times from September 1996 to
November 1999 as listed in Table 1. The two older observations (1996 
and 1997) were performed with all five PCA units operating, while in the
1999 observation  only three units were on. Standard selection
criteria were applied to the data excluding time intervals
corresponding to South Atlantic Anomaly passage, Earth's limb
lower than $10^\circ$ and angular distance between the source and
the satellite pointing direction larger than 0\fdeg02. We used
only data obtained with the PCA (Jahoda et al. 1996) accumulated
in ``Good Xenon" telemetry mode, time tagged with a 1 $\mu$s
accuracy with respect to the spacecraft clock and with absolute
time accuracy of 5--8 $\mu$s with respect to the UTC (Rots et al.
1998). Data from the first detector layer only are selected for
the analysis to increase the signal-to-noise ratio.

BeppoSAX observed the source on 17 March 
2000 for 196 ks elapsed time. In our analysis we consider only
data from the MECS (1.6--10 keV; Boella et al. 1997) whose total
exposure was 96.9 ks. MECS events were extracted from a circular 
region of 2' radius which maximize the signal-to-noise 
ratio of the pulsed component. 

In addition the BSAX and RXTE data, we used the pulsar spectrum from Becker
et al. (2003) which was taken with Chandra between July and September 2002.
For observational and data analysis details see Becker et al. (2003). 

The observing and  exposure times  of each observation are shown in
Table 1.

\begin{table}
\label{tab1} \caption{Observation log.}
\begin{tabular}{lccc}
\hline Instrument(ObsId)  & \multicolumn{2}{c}{Starting time } &
Exposure
\\ &     &(MJD) & (ks) \\
\hline
RXTE/PCA(P10421)&   16~Sep~1996  &50342.26 & 6.4 \\
 RXTE/PCA(P20159)&   10~Feb~1997  &50489.87 & 97.3 \\
 RXTE/PCA(P40090)&   12~Nov~1999  &51494.25 & 38.7 \\
 BSAX/MECS       &   17~Mar~2000  &51620.57 & 96.9 \\
 Chandra/ACIS    &   ~4~Jul~2002  &52459    &12.7 \\
 Chandra/ACIS    &   ~8~Aug~2002  &52494    &13.5 \\
 Chandra/ACIS    &   ~9~Sep~2002  &52526    &11.4 \\
\hline
\end{tabular}
\end{table}


\section{Timing analysis}

Arrival times were converted to the Solar System Barycenter using
the (J2000) pulsar coordinates given by 
Rutledge et al. (2003) (see Table 2) and the JPL2000 ephemeris 
(DE2000; Standish 1982).
Search for the period was performed with the folding technique in a
small frequency range around the value expected from the
Nan\c{c}ay radio ephemeris reported in Table 2 (Rutledge et al.
2003).

\begin{table}
\label{tab1} \caption{Radio ephemeris of \PSR (Rutledge et al.
2003).}
\begin{center}
\begin{tabular}{ll}
\hline Parameter & Value
\\
\hline
 Right Ascension (J2000) & $18^{\rm h}$ $24^{\rm m}$ $32.0^{\rm s}$    \\
 Declination (J2000) & $-24\degr$ $52\arcmin$ $10\farcs7$   \\
 Validity Epoch (MJD) & 50351--52610   \\
 DM (pc cm$^{-3}$)   & 119.873   \\
$\nu$ (Hz) & 327.40564101150(1) \\
 $\dot{\nu}(10^{-12}$ Hz s$^{-1})$& $-0.1735080(1)$\\
 $\ddot{\nu}(10^{-24}$ Hz s$^{-2})$& 0.66 \\
 $t_{0}$ (MJD) & 51468.0 \\
 \hline
\end{tabular}
\end{center}
\end{table}

The best frequency of each RXTE observation was computed fitting the
peak of the $\chi^{2}$ distribution with a gaussian and it was found
compatible with the expected value from radio measurements. 
Errors at $1 \sigma$ level were derived by the frequency interval 
corresponding to a unit decrement with respect to the maximum in the 
$\chi^{2}$ curve (${\rm err}=\nu(\chi^{2}_{max})-\nu(\chi^{2}_{max}-1)$). 
Table 3 shows the detected frequency for the entire data set at 
the reference time assumed equal to the central time of each observation; 
the values in parenthesis are the statistical uncertainty in the last
digit corresponding to one gaussian standard deviation.

\begin{table}
\label{tab1} \caption{Detected frequencies.}
\begin{center}
\begin{tabular}{lcl}
\hline Instrument  & Reference Time &
\multicolumn{1}{c}{Frequency}
\\
                   & (MJD)            & (Hz) \\
\hline
 RXTE/PCA&   50342.31617688   & 327.4056574(8) \\
 RXTE/PCA&   50490.95766180   & 327.40565566(1) \\
 RXTE/PCA&            51494.73885753   & 327.40564068(5)  \\
 BSAX/MECS$^{*}$  &   51621.70908612 & 327.4056429(4) \\
 \hline \\
\end{tabular}
\end{center}
$^{*}$Quoted errors are the statistical ones. A further
systematic error of $-3.2\times10^{-6}$ Hz must be considered
for this observation (see text).
\end{table}

\begin{figure}
\label{fig1} \centerline{ \vbox{
\psfig{figure=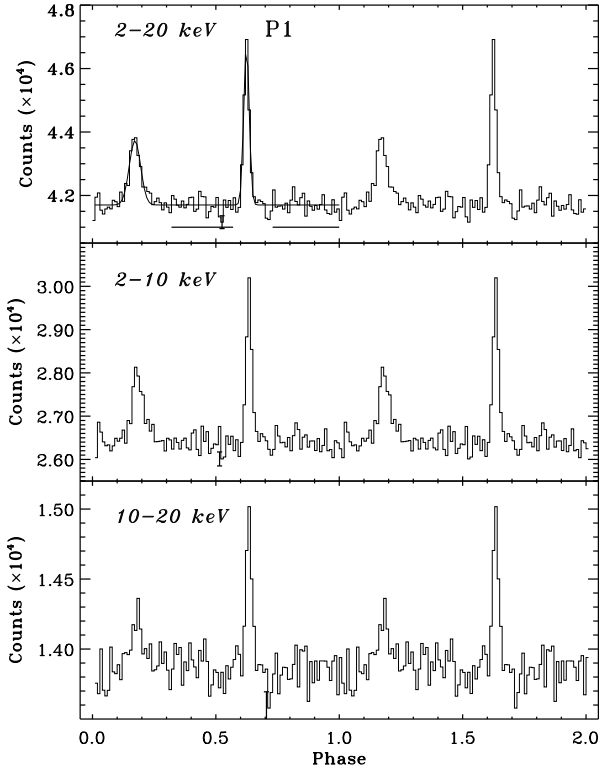,width=8cm,angle=0,clip=} }}
\caption{ PSR~B1821$-$24 phase histogram obtained with RXTE/PCA: in
 the top panel the 2--20 keV light curve is shown with the two Gaussian
 fitting model.
The two partial light curves 2--10  and 10--20 keV are shown in the middle 
and top panel, respectively. The off-pulse region used to extract
the spectrum are indicated in the top panel.}
\end{figure}

A cumulative RXTE light curve has been generated at the radio 
ephemeris adding the light curves of each observation. 
 The highest significance of 
the pulsed emission is reached in the energy interval 2--20 keV 
(PHA 5--54) and the resulting X-ray pulse profile is shown in the 
top panel of Fig. 1. Two peaks are evident in this light curve: a sharp
peak at phase $\Phi=0.62$ and a wider and lower peak at phase
$\Phi=0.18$. In this  paper, we use the nomenclature of Backer \&
Sallmen (1997) referring the peak at phase 0.62 as P1$_{\rm X}$ and the other as 
P2$_{\rm X}$. The
statistical significance is $19\sigma$ and $10\sigma$ for the 
first and second peak, respectively. The entire pulse profile
can be modeled by two Gaussian curves, one for each peak, plus a
constant level.  The statistical agreement is good and the Gaussian
parameters can be used to evaluate some interesting parameters:
the phase separation between the two peaks is $0.452\pm0.002$
measured from P2$_{\rm X}$ to P1$_{\rm X}$; 
the best fit parameters for P1$_{\rm X}$ and P2$_{\rm X}$
for the FWHM are $0.0259\pm0.0011$ and
$0.054\pm0.004$ for P1$_{\rm X}$ and P2$_{\rm X}$ respectively, and the normalization are
$(2.03 \pm 0.13) \times 10^3$   and $(5.08 \pm 0.19) \times 10 ^3$,
respectively. Two other pulse profiles
in narrower energy ranges (2--10 keV and 10--20 keV) are shown 
in the other panels of Fig. 1. 
To search for possible morphology changes, we computed the P2$_{\rm X}$/P1$_{\rm X}$ 
ratios in different energy bands.    
This calculation was performed from the total counts in the phase
intervals 0.60--0.65 and 0.13--0.24  for P1$_{\rm X}$ and P2$_{\rm X}$,
respectively, after the subtraction of the off-pulse level, 
estimated in the phase intervals 0.32--0.57 and 0.73--1.00  (see Fig. 1).  
The P2$_{\rm X}$/P1$_{\rm X}$ values of the two ratios were $0.80\pm0.06$ and $0.65\pm0.14$ 
in the 2--10 keV and 10--20 keV band respectively, indicating only a marginal 
variation at $1.5\sigma$ significance.   
We verified these results from the integrated Gaussian profiles, with the  
best fit parameters given above, and found an agreement within one standard deviation.

BeppoSAX data were analyzed with the same method and the 
resulting $\chi^{2}$ distribution of the MECS events showed 
a discrepancy between the measured frequency $\nu_{\rm X}$ 
and the expected one from radio ephemeris $\nu_{\rm R}$, 
$\nu_{\rm R}-\nu_{\rm X}=-3.2\times10^{-6}$ Hz. 
This difference is due to a systematic of the onboard clock 
present in the BeppoSAX observations after January 2000 
(Nicastro et al. 2002). 
For this reason, in the production of MECS light curves we 
used the proper ephemeris $\nu_{\rm X}$. 
The resulting 1.6--10 keV  profile  is shown in Fig. 2: 
two peaks with phase separation $0.44\pm0.02$ are evident. 
The profile is different from 
the RXTE one because  the worse BeppoSAX 
timing accuracy ($\sim 200$ $\mu$s; Mineo et al. 2000) widens the two peaks. 
We arbitrarily associated the higher and narrower peak to P1$_{\rm X}$ 
because the BeppoSAX accuracy does not allow the evaluation 
of the absolute phase.

The location of PSR~B1821$-$24 near the core of the globular cluster
makes difficut the evaluation of the pulsed fraction because of the 
contamination from these sources not resolved by the 
large MECS point spread function. Their contribution
in the extraction region was computed using  
the results  published in Becker et al. (2003) and subtracted
from the light curve. The resulting  signal
is consistent with being 100\% pulsed 
with a $2\sigma$ upper limit on the unpulsed emission of 20\%
in agreement with the Rutledge et al. (2003) results.

\begin{figure}
\label{fig2} \centerline{ \vbox{
\psfig{figure=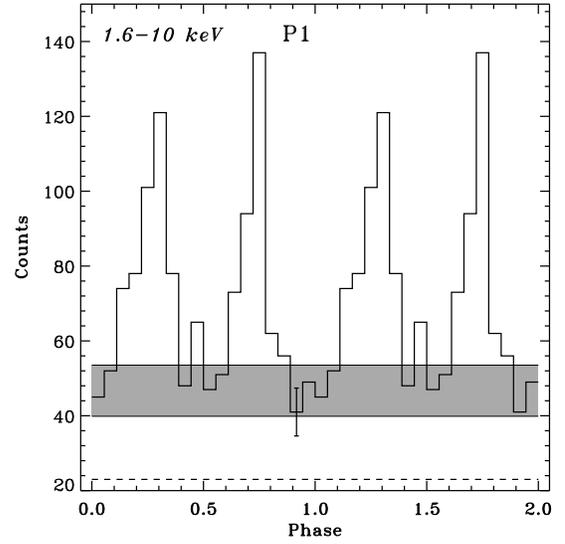,width=8cm,angle=0,clip=} }}
\caption{ PSR~B1821$-$24 phase histogram obtained by the BSAX/MECS
in the energy range 1.6--10 keV. The grey band indicates the DC level
and the dashed line the background.}
\end{figure}

The occurrence of giant pulses (GPs) at the same phase of the X-ray
peak P1$_{\rm X}$ suggests the possibility that energetic photons are emitted
during the same events in which the radio GPs are originated. 
To verify this hypothesis, we searched for a bunching of the X-ray 
photons, possibly with a rate similar to that of GPs. 
According to Romani \& Johnston (2001) the GP rate from \PSR is
$\sim 5$--6 events per hour, and therefore, if the X-ray flux
is more intense in these occasions, one can expect to observe  few 
pulses with a number of photons largely in excess to the
poissonian fluctuations. During the RXTE exposure we should expect
about 200 possible X-ray flares correlated with giant pulses.
In the search of these events, we applied the same method used in 
the analysis of the MSP PSR B1937+21 (Cusumano et al. 2003) and produced 
an X-ray light curve selecting only the events within a narrow 
 phase interval of width $\Delta\Phi=0.06$ (180 $\mu$s) centered at P1$_{\rm X}$ 
and studied the frequency distribution of these events.
This width has been chosen because it is comparable 
to the full width half maximum of P1$_{\rm X}$ and of the giant pulse 
phase window. 
Since the dead time of the PCA is about 10 $\mu$s, the maximum 
content of a bin in the presence of an X-ray flare cannot exceed 18-19 counts.
The
resulting distribution deviates from the Poisson law because
we found a number of bins having more than 2 events much higher than
expected. 
In particular, we obtained 1312 bins with 2 counts,
16 bins with 3 counts, 3 bins with 4 counts and 1 bin with 5 counts. 
However, comparing this distribution with others
from several phase intervals in the off-pulse region, we found a similar 
number of bins with more than two events. 
Furthermore, comparable excesses were also found in the analysis of PSR B1937+21
(Cusumano et al. 2003), and therefore they seems related to instrumental
timing performance rather than some other origin, like the variation of 
sources in the field of M28 (Becker et al. 2003).
We conclude that there is no evidence for a bunching of the 
X-ray emission.

\begin{figure}
\label{fig3} \centerline{ \vbox{
\psfig{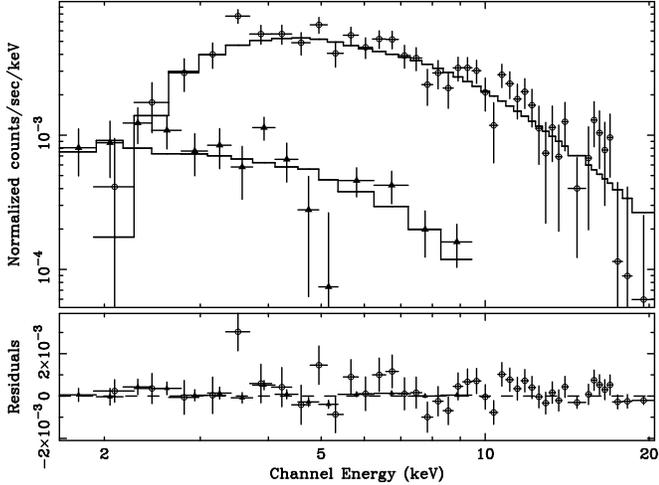}
 }}
\caption{PSR~B1821$-$24 pulsed spectrum with the best fit power-law 
model (top panel) and the relative residuals (bottom panel). 
Triangles indicate the MECS points and stars indicate RXTE data.}
\end{figure}

\section{Spectral analysis}
Pulse phase histograms were accumulated  for each unit of RXTE/PCA
independently for the 256 PHA channels. The same procedure was
applied to MECS data accumulating a phase resolved histogram
for each energy channel.

RXTE pectra of the two peaks were accumulated in the
phase intervals 0.60--0.65 and 0.13--0.24  for P1$_{\rm X}$ and P2$_{\rm X}$
respectively; off-pulse counts were extracted from the phase
intervals 0.32--0.57 and 0.73--1.00 (see Fig. 1). The total
pulsed spectrum was obtained from the sum of the two peaks.
Energy channels were rebinned in order to have at least 30 counts 
per bin and make the $\chi^{2}$ statistics available.
RXTE spectra were combined by summing the individual units and
assigning a total exposure time equal to the sum of the individual
exposures. 
RXTE response matrix were derived for each PCA unit and summed together 
weighting with the background subtracted counts of the correspondent 
PCU's phase histogram. 

A similar procedure was also followed for the MECS spectra.
In this case however, because of the broad peak profiles (see Fig. 2), 
the off-pulse was estimated from the interval 0.40--0.60.

RXTE (energy range 2--20 keV) and  MECS (energy range 1.6--10 keV)
spectra were simultaneously fitted  fixing the intercalibration factor
to the value obtained modelling the two spectra in the common range
2--10 keV with a power law. The value $1.12\pm0.05$ is in agreement
within two standard deviation 
with the values expected from the PCA-MECS intercalibration (Kuulkers et al.
2003).

Pulsed spectra were fitted with an absorbed power law with the 
galactic column density  fixed to the values
$N_{\rm H}=1.6\times10^{21}$ cm$^{-2}$ obtained by Chandra,
in agreement with that computed from 
the reddening $E({\rm B-V})$ of the globular cluster (Becker et al. 2003).

The  average pulsed spectral index was
$\alpha=1.30_{+0.12}^{-0.05}$ and the unabsorbed (2--10 keV)
pulsed X-ray flux was $3.52\times10^{-13}$ erg cm$^{-2}$ s$^{-1}$
corresponding to a uniform luminosity of $L_{\rm X} = 1.3\times 10^{33}
\; (d/5.5\;{\rm kpc})^2$ erg s$^{-1}$. 
This value agrees very well with the luminosity estimated by Saito et
 al. (1997) after the correction due to the small difference in the distance.
The ratio of the observed (2--10 keV) luminosity  with the spin-down energy loss is
$$ \frac{L_{\rm x}}{|\dot{E}_r|} =  5.8 \times 10^{-4}
\; \frac{(d/5.5\;{\rm kpc})^2}{I_{45}}  $$
where $I_{45}$ is the moment of inertia in 10$^{45}$ gr cm$^{2}$. It 
lies very nicely on the correlation between  X-ray and  spin-down
luminosity given by Becker \& Tr\"umper  (1999) and Possenti et al. (2002).
As already noticed by Becker et al. (2003), the value of $\dot{E}_r$ is
essentially intrinsic because the line-of-sight
projection of the acceleration in the cluster's gravitational potential
can affect  $\dot\nu$  by  an amount less
than 10\%.

The pulsed spectra with the fitting model (top panel) and the relative 
residuals (bottom panel) are shown in Fig. 3.
Only one bin around 3 keV shows a significant excess but it is too narrow
to be compatible with the instrumental resolution and cannot be due
to a true spectral feature. Moreover, adding a Gaussian line at 3.3 keV 
and having a width of 0.8 keV we obtained 
a $2\sigma$ upper limit of
$6.8\times10^{-6}$ photons cm$^{-2}$ s$^{-1}$,  which means that neither the
RXTE nor the BSAX data are sensitive enough to confirm this spectral feature
seen in the Chandra data.
Furthermore, a black body spectral distribution did not 
fit the average pulsed spectra giving an unacceptable reduced 
$\chi^2_{\rm red}=2.1$ (56 d.o.f.).

Fitting a power law spectrum to the two peaks separately, a marginal
difference of the spectral indices has been detected with
$\alpha=1.17_{-0.13}^{+0.06}$ and $\alpha=1.38_{-0.09}^{+0.18}$ 
for P1$_{\rm X}$ and P2$_{\rm X}$, respectively. Table 3 reports the best fit 
parameters for the two peaks fitted independently and for the total
pulsed emission (P1$_{\rm X}$+P2$_{\rm X}$).

To improve the fit accuracy, we performed the spectral analysis 
of the total signal from \PSR considering also Chandra/ACIS data. 
This spectrum is relative to the total pulsed emission and includes 
an unpulsed component that, according Rutledge et al. (2003), can
account for 15\% of the total emission; presently there is no spectral
information about this component. Taking into account these remarks, we fitted 
the pulsed RXTE and MECS spectra together with the Chandra total one  
using a single power law. The Chandra/ACIS intercalibration factor  
was determined with respect to RXTE by fitting the events in the common 
energy ranges and resulted equal to $0.91\pm 0.07$. The spectral index 
was found fully compatible with the previous analysis but with a much smaller
uncertainty.
The subtraction of the RXTE pulsed spectrum, extrapolated 
at energies below 2 keV, from the Chandra data did not give any 
significant low energy residuals, indicating 
that the DC component reported by Rutledge et al. (2003),
if detectable, does not have a spectrum much softer than the
pulsed one. An estimate of significant constraints on the spectral shape is 
practically not possible because of the large 
uncertainty on the intercalibration factors.

\begin{table}
\label{tab1} \caption{Best fit parameters of the power-law model.}
\begin{tabular}{lccc}
\hline
Component & \multicolumn{1}{c}{$\alpha$}& \multicolumn{1}{c}{$F$(2--10~keV)$^1$}
 & \multicolumn{1}{c}{$\chi^2_{\rm r}$(d.o.f.)} \\
\hline
&                       &               &         \\
P1$_{\rm X}$&$1.17_{-0.13}^{+0.06}$ & $1.82\pm0.24$ & 0.93(56)  \\
&                       &               &         \\
P2$_{\rm X}$&$1.38_{-0.09}^{+0.18}$ & $1.67\pm0.21$ & 1.22(56)  \\
&                       &               &         \\
P1$_{\rm X}$+P2$_{\rm X}$ &$1.30_{-0.05}^{+0.11}$ & $3.52\pm0.33$ & 1.07(56)  \\
     &                       &               &         \\
&                       &               &         \\
P1$_{\rm X}$+P2$_{\rm X}$+(DC)$^2$ & $1.30_{-0.02}^{+0.05}$ & $3.52\pm0.33$ & 0.97(91) \\
&                       &               &         \\
\hline
\end{tabular}

$^1$ in units of 10$^{-13}$ erg cm$^{2}$ s$^{-1}$ \\
$^2$ MECS, RXTE and Chandra (total signal) data
\end{table}

\section{Discussion}
Our analysis of the available X-ray observations of \PSR performed with
 RXTE, BeppoSAX and Chandra confirms the known picture on this source.
In particular, the joint analysis of the whole data set provided
an accurate estimate of the phase averaged photon index which
resulted $1.30_{+0.05}^{-0.02}$;  moreover, RXTE and BeppoSAX
gave an indication of a spectral difference between the two X-ray peaks.
\PSR and PSR~B1937+21 are the only two MSPs whose radio emission
shows giant pulses concentrated in the same narrow phase window
of X-ray pulses (Kinkhabwala \& Thorsett 2000, Romani \& Johnston 2001,
Cusumano et al. 2003).
It is unlikely that this phase coincidence would be casual and therefore
it can be an indication that radio GPs and X-rays are emitted in the
same region of the magnetosphere.
For this reason we focused our attention to verify if the X-ray
emission is occasionally enhanced with a rate comparable to that
of GPs. A statistical test on the distribution of the number of X-ray
events gave a negative result, indicating that, differently from
GPs, X-rays photons are emitted in a steady process.
Furthermore, the fact that the phase distances between normal radio
and X-ray pulses are about the same in these two MSPs indicates
that their respective origin regions do not have a large spatial
separation.

It is possible that the high-energy  emission is enhanced in coincidence
with radio GPs. This effect has been found in the Crab pulsar whose
optical pulses intensity simultaneous to GPs increases on average
of 3\% with respect to the normal radio pulses (Shearer et al. 2003).
This result has not been verified at $\gamma$-ray energies though the present
limit cannot exclude it (Lundgren et al. 1995).
As pointed 
out by Shearer et al. (2003) an increase of the pair plasma
density of the same order of the optical intensity cannot completely
explain the very large flux of radio GPs, which in Crab are orders of
magnitude stronger than normal pulses.

In the case of \PSR we can derive a crude upper limit of the pulsed signal fraction
emitted in correlation with the GPs.   
Assuming that $\sim10\%$ of the total pulsed counts are concentrated 
within the GP phase window, we should expect several bins with 6 or more 
events in a single pulse; but they are not observed indicating that
the fraction must be less than the above value.
The same limit applies to the total number of emitting particles, whereas
it does not hold for the local fluctuations of their density within
a small region of the magnetosphere.

The actual site in the magnetosphere where pulses originate is uncertain.
In some models the phase separations close to 0.5 has been considered
as an indication that pulses originate from the polar regions of an
orthogonal rotator (see e.g. Ruderman 2003).
Electrons should then be accelerated in the polar gap and emit
coherent and incoherent radiation from this site.
Romani \& Johnston (2001) proposed for \PSR a different scenario in
which GPs and X rays are emitted in the outer gaps, where a high local
magnetic field should enhance the synchrotron emissivity and the production
rate of secondary e$^+$e$^-$ pairs. X-ray pulses are then emitted
by this dense pair plasma that would also produce instabilities
which in turn increase the particles coherence and hence
the emission of radio GPs.
Pulse profiles for the outer gap geometry have been calculated by
Wang et al. (2002), who obtained narrow radio and X-ray peaks with
a phase separation close to 0.5.  Whether these  models are appropriate
to describe the observed scenario can be decided only when more pulsars
showing the same behaviour are detected. 

\begin{acknowledgements}
This work has been partially supported by INAF (Istituto 
Nazionale di Astrofisica).
\end{acknowledgements}

\end{document}